\documentclass[preprint,12pt]{elsarticle}

\usepackage{amssymb}
\usepackage{amsmath,amssymb,amscd,amsbsy,amsgen,amsopn,amstext,amsxtra}
\usepackage[mathscr]{eucal}

\newcommand{\s}{\:\!}
\newcommand{\m}{\;\!}
\newcommand{\ket}{\rangle}
\newcommand{\bra}{\langle}

\newcommand{\dket}{\rangle\!\rangle}
\newcommand{\dbra}{\langle\!\langle}
\newcommand{\dcket}{)\!)}
\newcommand{\dcbra}{(\!(}

\newcommand{\1}{\mbox{1}\hspace{-0.25em}\mbox{l}}

\begin{document}

\begin{frontmatter}


\title{Two quantization approaches to the Bateman oscillator model}




\author[Deg]{Shinichi Deguchi}
\ead{deguchi@phys.cst.nihon-u.ac.jp}

\address[Deg]{Institute of Quantum Science, College of Science and Technology, 
Nihon University, Chiyoda-ku, Tokyo 101-8308, Japan}

\author[Fuj]{Yuki Fujiwara\corref{mycorrespondingauthor}}
\cortext[mycorrespondingauthor]{Corresponding author.}
\ead{yfujiwara@phys.cst.nihon-u.ac.jp}

\address[Fuj]{Department of Quantum Science and Technology, Graduate School of Science and Technology, 
Nihon University, Chiyoda-ku, Tokyo 101-8308, Japan}

\author[Nak]{Kunihiko Nakano}
\ead{knakano@sic.shibaura-it.ac.jp}

\address[Nak]{Junior and Senior High School, Shibaura Institute of Technology, Koto-ku, Tokyo 135-8139, Japan}

\begin{abstract}
We consider two quantization approaches to the Bateman oscillator model. 
One is Feshbach-Tikochinsky's quantization approach reformulated concisely without invoking
the ${\mathit{SU}(1,1)}$ Lie algebra, and the other is the imaginary-scaling quantization approach 
developed originally for the Pais-Uhlenbeck oscillator model. 
The latter approach overcomes the problem of unbounded-below energy spectrum 
that is encountered in the former approach. 
In both the approaches, the positive-definiteness of the squared-norms of the Hamiltonian eigenvectors is ensured. 
Unlike Feshbach-Tikochinsky's quantization approach, 
the imaginary-scaling quantization approach allows to have stable states in addition to 
decaying and growing states. 

\end{abstract}


\end{frontmatter}





\numberwithin{equation}{section}
\section{Introduction}

The Bateman oscillator model \cite{Bateman}, or simply the Bateman model, has repeatedly been investigated 
as a Lagrangian model for the damped harmonic oscillator since Bateman presented 
the model about 90 years ago 
\cite{MorFes, Dekker, Razavy, FesTik, CRV, SVW, BGPV, BlaJiz, ChrJur, Chruscinski, BanMuk}.  
The Bateman Lagrangian, which governs the Bateman model, in actuality describes 
a doubled system consisting of the (uncoupled) damped and amplified harmonic oscillators. 
Nevertheless, the Bateman model is widely recognized as a standard model for the damped harmonic oscillator, 
because the Bateman Lagrangian yields the correct equation of motion of the damped harmonic oscillator  
and has the desirable property that the Lagrangian itself does not explicitly depend on time.

Canonical quantization of the Bateman model was first performed by Feshbach and Tikochinsky 
with the aid of the representation theory of the ${\mathit{SU}(1,1)}$ Lie algebra \cite{FesTik}. 
They obtained the eigenvalues of the Hamiltonian operator of the Bateman model and their corresponding eigenvectors. 
These eigenvalues are necessarily complex numbers, and hence the corresponding eigenstates 
(in the Schr\"{o}dinger picture) turn out to be either decaying or growing states. 
Also, it is seen that the real parts of the Hamiltonian eigenvalues, which can be identified as 
possible values of energy of the system, are unbounded from below. 
From a purely dynamical point of view, this will cause the problem 
of dynamical instability of the system if interactions are turned on. 
(Applying the framework of thermo field dynamics (TFD) \cite{TakUme, Umezawa} to quantizing the Bateman model 
may bypass this problem \cite{CRV, SVW}.)
After Feshbach and Tikochinsky performed the quantization of the Bateman model, 
their results have been reconsidered in some different contexts 
\cite{CRV, SVW, BGPV, BlaJiz, ChrJur, Chruscinski}.   
However, it seems that the problem of unbounded-below energy spectrum 
has not been raised precisely and has not been resolved yet.\footnote{~Recently, 
quantization of the Bateman model has been studied in connection with 
a noncommutative space \cite{PNC}. 
For other recent studies concerning quantization of the Bateman model, see, e.g., Refs. \cite{Bagarello, GLAC}. 
The contents in these studies are not directly related to those treated in the present paper.}

A similar problem is encountered in the canonical quantization of the Pais-Uhlenbeck oscillator model \cite{PaiUhl}. 
Since the Lagrangian of this model contains the second order time-derivative of a coordinate variable 
in a non-degenerate manner, the corresponding classical Hamiltonian turns out to be unbounded from below 
in accordance with the Ostrogradsky theorem \cite{Ostrogradsky, Woodard}.  
This undesirable situation is inherited by the standard canonical quantization of the Pais-Uhlenbeck model, 
leading to the problem of unbounded-below energy spectrum. 
In order to overcome this problem, with ensuring the positive-definiteness of the squared-norms of 
the corresponding eigenvectors, Bender and Mannheim proposed an alternative quantization scheme involving 
the imaginary scaling of position and momentum operators \cite{BenMan}.  
Subsequently, Mostafazadeh explored mathematical aspects of this quantization scheme and 
called it the {\em imaginary-scaling quantization} \cite{Mostafazadeh}. 
This scheme indeed gives the bounded-below energy spectrum 
having no corresponding eigenvectors of negative squared-norm.

In this paper, we apply the imaginary-scaling quantization scheme to the Bateman model 
to obtain the Hamiltonian eigenvalues whose real parts are bounded from below. 
Of course, the positive-definiteness of the squared-norms of the corresponding eigenvectors 
is precisely taken into account. 
Before proceeding to the imaginary-scaling quantization approach to the Bateman model, 
we first attempt to concisely reformulate Feshbach-Tikochinsky's quantization approach 
by exploiting a pseudo Bogoliubov transformation. 
It will be seen that our method does not invoke the ${\mathit{SU}(1,1)}$ Lie algebra 
and hence is simpler and less complicated than 
the original Feshbach-Tikochinsky's method \cite{FesTik, CRV, ChrJur, Razavy}.  
This is a point that we would like to stress here. 
After the reformulation of Feshbach-Tikochinsky's approach,  
we indeed develop the imaginary-scaling quantization approach to the Bateman model 
by exploiting the combination of an imaginary-scaling transformation and a homogeneous transformation. 
We will see that the two quantization approaches are realized on an equal footing 
on the basis of different transformations of the annihilation and creation operators.

This paper is organized as follows: 
Section 2 provides a brief review of the Bateman model, including a preparation for the two quantization approaches. 
In Section 3, we reformulate Feshbach-Tikochinsky's quantization approach in a concise manner, 
and in Section 4, we study the imaginary-scaling quantization approach to the Bateman model. 
Section 5 is devoted to a summary and discussion.

\numberwithin{equation}{section}
\section{Bateman model and its canonical quantization}

In this section, we briefly review the Bateman model and its quantum-mechanical setup. 

The Bateman model is defined by the Lagrangian \cite{Bateman} 
\begin{align}
L=m\dot{x}\dot{y}+\frac{\gamma}{2}(x\dot{y}-\dot{x}y)-kxy \,, 
\label{2.1}
\end{align}
where $x=x(t)$ and $y=y(t)$ are real coordinate variables, being functions of time $t$, 
and $m$, $\gamma$, and $k$ are real positive constants.\footnote{~The Lagrangian (\ref{2.1})  
is invariant under the transformation 
$(x, y, \gamma) \rightarrow (^{\s}y, x, -\gamma)$. 
Accordingly, Eqs. (\ref{2.2}) and (\ref{2.3}) are related to each other by this transformation.}
A dot over a variable denotes its derivative with respect to $t$. 
From this Lagrangian, the Euler-Lagrange equation for $y$ is derived as  
\begin{align}
m\ddot{x}+\gamma \dot{x} +kx=0 \,,
\label{2.2}
\end{align}
and similarly, the Euler-Lagrange equation for $x$ is derived as 
\begin{align}
m\ddot{y}-\gamma \dot{y} +ky=0 \,.
\label{2.3}
\end{align}
Equation (\ref{2.2}) is precisely the classical equation of motion 
for the damped harmonic oscillator 
of mass $m$, spring constant $k$, and damping constant $\gamma$. 
Equation (\ref{2.3}) is the classical equation of motion for 
the amplified harmonic oscillator whose amplitude exponentially grows with time 
while the amplitude of the damped harmonic oscillator exponentially decays with time. 
We thus see that the Bateman Lagrangian (\ref{2.1}) describes a doubled system 
consisting of the (uncoupled) damped and amplified harmonic oscillators.

Let us introduce the new variables \cite{BGPV, BanMuk} 
\begin{align}
x_{1}:=\frac{1}{\sqrt{2}} (x+y) \,, 
\quad \;\, 
x_{2}:=\frac{1} {\sqrt{2}} (x-y) \,,
\label{2.4}
\end{align}
with which the Lagrangian (\ref{2.1}) can be written as 
\begin{align}
L=\frac{m}{2}\Big( \dot{x}_{1}^{2} -\dot{x}_{2}^{2} \Big)
-\frac{\gamma}{2} \big( x_{1} \dot{x}_{2} -\dot{x}_{1} x_{2} \big) 
-\frac{k}{2} \Big( x_{1}^{2} -x_{2}^{2} \Big) \m.
\label{2.5}
\end{align}
The momenta conjugate to $x_{1}$ and $x_{2}$ are found to be 
\begin{align}
p_{1}:=\frac{\partial L}{\partial \dot{x}_{1}} =m\dot{x}_{1}+\frac{\gamma}{2} x_{2} \m, 
\quad \;\, 
p_{2}:=\frac{\partial L}{\partial \dot{x}_{2}} =-m\dot{x}_{2}-\frac{\gamma}{2} x_{1} \m. 
\label{2.6}
\end{align}
The Hamiltonian is obtained by the Legendre transformation of $L$ as follows: 
\begin{align}
H&:=p_{1}\dot{x}_{1}+p_{2}\dot{x}_{2}-L
\notag
\\
&\;=\left( \frac{1}{2m} p_{1}^{2} +\frac{1}{2} m\omega^2 x_{1}^{2} \right)
-\left( \frac{1}{2m} p_{2}^{2} +\frac{1}{2} m\omega^2 x_{2}^{2} \right) 
-\frac{\gamma}{2m} (x_{1} p_{2}+x_{2} p_{1} ) \,, 
\label{2.7}
\end{align}
where 
\begin{align}
\omega:=\sqrt{ \frac{k}{m}-\frac{\gamma^2}{4m^2} }\,. 
\label{2.8}
\end{align}
In this paper, we treat only the underdamped-underamplified case 
by assuming that $\omega$ is real and positive.

Now, regarding the canonical variables $x_{i}$ and $p_{i}$ $(^{\s} i=1, 2^{\s}$) as 
Hermitian operators satisfying ${x_{i}^{\dagger}=x_{i}}$ and ${p_{i}^{\dagger}=p_{i}}$, 
we perform the canonical quantization of the Bateman model by imposing 
the commutation relations 
\begin{align}
\left[\s x_{i}\s,\s p_{j} \right]=i\hbar \delta_{ij} \1 \quad (\s i, j=1, 2 \s) \,, 
\quad\: \mbox{all others}=0 \,,  
\label{2,9}
\end{align}
where $\1$ denotes the identity operator. 
In terms of the operators 
\begin{subequations}
\label{2.10}
\begin{align}
a_{i} &:=\sqrt{\frac{m\omega}{2\hbar}}\m x_{i} +i \sqrt{\frac{1}{2\hbar m\omega}} \m p_{i} \,,
\label{2.10a}
\\[3pt]
a_{i}^{\dagger} &:=\sqrt{\frac{m\omega}{2\hbar}}\;\! x_{i} -i \sqrt{\frac{1}{2\hbar m\omega}} \;\! p_{i} \,, 
\label{2.10b}
\end{align}
\end{subequations}
which satisfy
\begin{align}
\left[\s a_{i}\s, \s a_{j}^{\dagger} \s \right]\!=\delta_{ij} \1 \,, 
\quad\: \mbox{all others}=0 \,, 
\label{2.11}
\end{align}
the Hamiltonian operator corresponding to the Hamiltonian (\ref{2.7}) can be expressed as 
\begin{align}
H=H_{0}+H_{1} \m,
\label{2.12}
\end{align}
with 
\begin{subequations}
\label{2.13}
\begin{align}
H_{0} &:=\hbar \omega \left( a_{1}^{\dagger} a_{1} -a_{2}^{\dagger} a_{2} \right) ,
\label{2.13a}
\\[2pt]
H_{1} &:=i \frac{\hbar\gamma}{2m}  \left( a_{1} a_{2} -a_{1}^{\dagger} a_{2}^{\dagger} \right) . 
\label{2.13b}
\end{align}
\end{subequations}
As can be readily seen, $H_{0}$ and $H_{1}$ are Hermitian 
(with respect to the $\dagger$-conjugation) and commute. 
Adopting the naive vacuum state vector $|^{\s} 0\ket$ specified by 
\begin{align}
a_{i}\s |\s0\ket=0 \,,
\label{2.14}
\end{align}
we can construct the Fock basis vectors 
\begin{align}
|^{\s} n_1, n_2 \ket :=\frac{1}{\sqrt{n_{1}! \, n_{2}!}} 
\Big(a_{1}^{\dagger} \Big)^{n_1} \Big(a_{2}^{\dagger} \Big)^{n_2} |\s0\ket 
\quad \,
(\s n_{i}=0, 1, 2, \ldots)\,. 
\label{2.15}
\end{align}
In this case, $a_{i}$ and $a_{i}^{\dagger}$ are identified as annihilation and creation operators, respectively. 
The dual forms of Eqs. (\ref{2.14}) and (\ref{2.15}) are given by
\begin{align}
\bra 0\s|\s a_{i}^{\dagger}=0 \,, 
\label{2.16} 
\end{align}
\vspace{-5mm}
\begin{align}
\bra n_1, n_2 |:=\frac{1}{\sqrt{n_{1}! \, n_{2}!}} 
\bra 0\s|(a_{1})^{n_1} (a_{2})^{n_2} \m.  
\label{2.17}
\end{align}
Using Eqs. (\ref{2.11}), (\ref{2.14}), and (\ref{2.16}), 
and imposing the normalization condition ${\bra 0\s|\s0\ket=1}$, 
we can show that 
\begin{align}
\bra m_1, m_2 \s |^{\s} n_1, n_2 \ket=\delta_{m_1 n_1} \delta_{m_2 n_2} \m.
\label{2.18}
\end{align}
Hence, it follows that the Fock basis vectors $|^{\s} n_1, n_2 \ket$ have the positive squared-norm 1, and 
the Fock space spanned by the orthonormal basis $\big\{^{\m} |^{\s} n_1, n_2 \ket \big\}$ 
is a positive-definite Hilbert space. 
In this space, the completeness relation of the orthonormal basis reads 
\begin{align}
\sum_{n_1=0}^{\infty} \sum_{n_2=0}^{\infty}  |^{\s} n_1, n_2 \ket \bra n_1, n_2 \s |=\1 \,. 
\label{2.19}
\end{align}
We see that the vectors $|^{\s} n_1, n_2 \ket$ are eigenvectors of $H_{0}$ 
but not eigenvectors of $H_{1}$, although $H_{0}$ and $H_{1}$ commute. 
In order to find the simultaneous eigenvectors of $H_{0}$ and $H_{1}$, 
which are of course eigenvectors of $H$, 
we consider invertible transformations in the next two sections.

\section{Feshbach-Tikochinsky's quantization approach revisited}

In this section, we perform a reformulation of the canonical quantization approach 
of Feshbach and Tikochinsky \cite{FesTik} 
without referring to the ${\mathit{SU}(1,1)}$ Lie algebra.

We first define the operators $\bar{a}_{i}$ and $\bar{a}_{i}^{\ddagger}$ by
\begin{align}
\bar{a}_{i}:=e^{\theta X} a_{i} \m e^{-\theta X} \s,
\quad\;\,
\bar{a}_{i}^{\ddagger}:=e^{\theta X} a_{i}^{\dagger} e^{-\theta X} \s,
\label{3.1}
\end{align}
where $\theta$ is a complex parameter, and $X$ is defined by
\begin{align}
X:=a_{1} a_{2} +a_{1}^{\dagger} a_{2}^{\dagger} \,.
\label{3.2}
\end{align}
It is obvious that ${X^{\dagger}=X}$. The unitarity of $e^{\theta X}$ and its associated property 
$(\bar{a}_{i})^{\dagger}=\bar{a}_{i}^{\ddagger}$ hold only when $\theta$ is purely imaginary. 
From Eq. ({2.11}), we see that 
\begin{align}
\left[\s \bar{a}_{i}\s, \s\bar{a}_{j}^{\ddagger} \s \right]\!=\delta_{ij} \1 \,, 
\quad\; \mbox{all others}=0 \,. 
\label{3.3}
\end{align}
Equation (\ref{3.1}) can be written as
\begin{subequations}
\label{3.4}
\begin{align}
\left( 
\begin{array}{c}
\bar{a}_{1} \\
\bar{a}_{2}^{\ddagger} \\ 
\end{array}
\right)
&=
\left(
\begin{array}{cc}
\cos\theta &\, -\sin\theta \\
\sin\theta &\, \cos\theta \\
\end{array}
\right) 
\left( 
\begin{array}{c}
a_{1} \\
a_{2}^{\dagger} \\ 
\end{array}
\right) \s,
\label{3.4a}
\\[5pt]
\left( 
\begin{array}{c}
\bar{a}_{1}^{\ddagger} \\
\bar{a}_{2} \\ 
\end{array}
\right)
&=
\left(
\begin{array}{cc}
\cos\theta &\, \sin\theta \\
-\sin\theta &\, \cos\theta \\
\end{array}
\right) 
\left( 
\begin{array}{c}
a_{1}^{\dagger} \\
a_{2} \\ 
\end{array}
\right) \s.
\label{3.4b}
\end{align}
\end{subequations}
The transformation 
$\big(a_{i}, a_{i}^{\dagger}\big) \rightarrow \big(\bar{a}_{i}, \bar{a}_{i}^{\ddagger} \big)$ 
looks like a Bogoliubov transformation, but actually it is not the case unless 
the parameter $\theta$ is purely imaginary. 
(If $\theta$ is purely imaginary, then $e^{\theta X}$ is unitary, and the transformation 
$\big(a_{i}, a_{i}^{\dagger}\big) \rightarrow \big(\bar{a}_{i}, \bar{a}_{i}^{\ddagger} \big)$ 
can be said to be a Bogoliubov transformation \cite{Bogoliubov, Umezawa}.)

Using Eqs. (\ref{3.4}) and (\ref{3.3}), we can express the operators $H_{0}$ and $H_{1}$ as follows: 
\begin{subequations}
\label{3.5}
\begin{align}
H_{0} &=\hbar \omega \left( \bar{a}_{1}^{\ddagger} \bar{a}_{1} -\bar{a}_{2}^{\ddagger} \bar{a}_{2} \right) ,
\label{3.5a}
\\[2pt]
H_{1} &=i \frac{\hbar\gamma}{2m}  
\left\{ \left( \bar{a}_{1} \bar{a}_{2} -\bar{a}_{1}^{\ddagger} \bar{a}_{2}^{\ddagger} \right) \cos(2\theta) 
+\left( \bar{a}_{1}^{\ddagger} \bar{a}_{1} +\bar{a}_{2}^{\ddagger} \bar{a}_{2} +\1 \right) \sin(2\theta) \right\} . 
\label{3.5b}
\end{align}
\end{subequations}
Incidentally, $X$ can be expressed as 
${X=\bar{a}_{1} \bar{a}_{2} +\bar{a}_{1}^{\ddagger} \bar{a}_{2}^{\ddagger}}$.  
Since our present purpose is to find the eigenvalues of $H$, we choose $\theta$ in such a way 
that $H_{1}$ takes the form of a linear combination of  
$\bar{a}_{1}^{\ddagger} \bar{a}_{1}$, $\bar{a}_{2}^{\ddagger} \bar{a}_{2}$, and $\1$. 
(The operator $H_{0}$ already takes the form of a  linear combination of  
$\bar{a}_{1}^{\ddagger} \bar{a}_{1}$ and $\bar{a}_{2}^{\ddagger} \bar{a}_{2}$.)  
Upon comparison with Feshbach-Tikochinsky's quantization approach,  we set ${\theta=\pm \pi/4}$. 
Then $H_{1}$ becomes  
\begin{align}
H_{1}^{(\pm)} &:=\pm i \frac{\hbar\gamma}{2m}  
\left( \bar{a}_{1}^{\ddagger} \bar{a}_{1} +\bar{a}_{2}^{\ddagger} \bar{a}_{2} +\1 \right) . 
\label{3.6}
\end{align}
The transformation 
$\big(a_{i}, a_{i}^{\dagger}\big) \rightarrow \big(\bar{a}_{i}, \bar{a}_{i}^{\ddagger} \big)$ with ${\theta=\pm \pi/4}$ 
is hereafter referred to as a {\em pseudo} Bogoliubov transformation,   
with the connotation that it is not unitary. 
Such a non-unitary transformation was also considered in 
Feshbach-Tikochinsky's quantization approach based on the ${\mathit{SU}(1,1)}$ Lie algebra. 
The Hermiticity of $H_{1}^{(\pm)}$ with respect to the $\ddagger$-conjugation, i.e., 
$\big(H_{1}^{(\pm)} \big)^{\ddagger}=H_{1}^{(\pm)}$ is valid under the conditions 
\begin{align}
i^{\s\ddagger}=-i\,, \quad\;\, \gamma^{\ddagger}=-\gamma \,.
\label{3.7}
\end{align}
Clearly, $H_{0}$ and $X$ are Hermitian with respect to the $\ddagger$-conjugation.

The Hamiltonian operator (\ref{2.12}) now reads $H^{(\pm)}=H_{0}+H_{1}^{(\pm)}$. 
With $H^{(\pm)}$, the Heisenberg equation for an implicitly time-dependent operator $A(t)$ reads 
$dA/dt=(i\hbar)^{-1} \big[^{\s}A, H^{(\pm)} \big]$. 
Using the commutation relations in Eq. (\ref{3.3}), 
we can solve the Heisenberg equations for $\bar{a}_{i}$ and $\bar{a}_{i}^{\ddagger}$    
and obtain   
\begin{subequations}
\label{3.8}
\begin{alignat}{3}
\bar{a}_{1}(t) &=\bar{a}_{1}(0)\s e^{(-i\omega \pm \lambda)\s t} \,, & \quad\;\, 
\bar{a}_{1}^{\ddagger}(t) &=\bar{a}_{1}^{\ddagger}(0) \s e^{-(-i\omega \pm \lambda)\s t} \,,
\label{3.8a}
\\[2pt]
\bar{a}_{2}(t) &=\bar{a}_{2}(0)\s e^{(i\omega \pm \lambda)\s t} \,, & \quad\;\,
\bar{a}_{2}^{\ddagger}(t) &=\bar{a}_{2}^{\ddagger}(0)\s e^{-(i\omega \pm \lambda)\s t} \,,
\label{3.8b}
\end{alignat}
\end{subequations}
where ${\lambda:=\gamma/2m}$. 
By virtue of the conditions in Eq. (\ref{3.7}), 
the $\ddagger$-conjugation relation  
${\big(\bar{a}_{i}(t)\big)^{\ddagger}=\bar{a}_{i}^{\ddagger}(t)}$ holds at an arbitrary time. 
As can be seen from Eq. (\ref{3.8}), the $\ddagger$-conjugation involves time reversal. 
This fact reminds us that in Feshbach-Tikochinsky's quantization approach, 
the time reverse, rather than the complex conjugate,  
is used to define an appropriate normalization integral for a wave function.  
It is evident that the Hamiltonian operator $H^{(\pm)}$ is independent of time.

Next we define the new vectors  
\begin{subequations}
\label{3.9}
\begin{align}
|\s0\dket &:=e^{\theta X} |\s0\ket \,, 
\label{3.9a}
\\[2pt]
\dbra 0\s| &:=\bra 0\s|\s e^{-\theta X} \s, 
\label{3.9b}
\end{align}
\end{subequations}
which satisfy 
\begin{align}
\bar{a}_{i}\s |\s0\dket=0 \,, 
\quad\;\,
\dbra 0\s|\s \bar{a}_{i}^{\ddagger}=0 
\label{3.10}
\end{align}
owing to Eqs. (\ref{2.14}) and (\ref{2.16}). 
Hence, ${|\s0\dket}$ and ${\dbra 0\s|}$ are established as the vacuum state vectors of the 
$\big(\bar{a}_{i}, \bar{a}_{i}^{\ddagger} \big)$-system, and $\bar{a}_{i}$ and $\bar{a}_{i}^{\ddagger}$  
turn out to be annihilation and creation operators, respectively. 
In this system, we can construct the Fock basis vectors and their dual vectors as follows: 
\begin{subequations}
\label{3.11}
\begin{align}
|^{\s} n_1, n_2 \dket &:=\frac{1}{\sqrt{n_{1}! \, n_{2}!}} 
\Big(\bar{a}_{1}^{\ddagger} \Big)^{n_1} \Big(\bar{a}_{2}^{\ddagger} \Big)^{n_2} |\s0\dket \m, 
\label{3.11a}
\\[2.5pt]
\dbra n_1, n_2 | &:=\frac{1}{\sqrt{n_{1}! \, n_{2}!}} 
\dbra 0\s|(\bar{a}_{1})^{n_1} (\bar{a}_{2})^{n_2} \s. 
\label{3.11b}
\end{align}
\end{subequations}
They are related to the old basis vectors in Eqs. (\ref{2.15}) and (\ref{2.17}) by 
\begin{subequations}
\label{3.12}
\begin{align}
|^{\s} n_1, n_2 \dket &=e^{\theta X} |^{\s} n_1, n_2 \ket \,, 
\label{3.12a}
\\[2pt]
\dbra n_1, n_2 | &=\bra n_1, n_2 |\s e^{-\theta X} \s. 
\label{3.12b}
\end{align}
\end{subequations}
By using Eq. (\ref{2.18}), it is easily shown that 
\begin{align}
\dbra m_1, m_2 \s |^{\s} n_1, n_2 \dket=\delta_{m_1 n_1} \delta_{m_2 n_2} \m.
\label{3.13}
\end{align}
Hence, it follows that the Fock basis vectors $|^{\s} n_1, n_2 \dket$ also have the positive squared-norm 1, and 
the Fock space spanned by the orthonormal basis $\big\{^{\m} |^{\s} n_1, n_2 \dket \big\}$ 
is a positive-definite Hilbert space. 
The completeness relation (\ref{2.19}) leads to 
\begin{align}
\sum_{n_1=0}^{\infty} \sum_{n_2=0}^{\infty}  |^{\s} n_1, n_2 \dket \dbra n_1, n_2 \s |=\1 \,.
\label{3.14}
\end{align}
We readily see that the vectors $|^{\s} n_1, n_2 \dket$ with ${\theta=\pm\pi/4}$ are simultaneous eigenvectors of 
$H_{0}$ and $H_{1}^{(\pm)}$ and satisfy the Hamiltonian eigenvalue equation 
\begin{align}
H^{(\pm)} |^{\s} n_1, n_2 \dket =h^{(\pm)}_{n_1,\m n_2} |^{\s} n_1, n_2 \dket 
\label{3.15}
\end{align}
with   
\begin{align}
h^{(\pm)}_{n_1,\m n_2} :=\hbar \omega (n_1 -n_2) 
\pm i\hbar \lambda (n_1 +n_2 +1) \,.
\label{3.16}
\end{align}
The Hamiltonian eigenvalues $h^{(\pm)}_{n_1,\m n_2}$ are identical to those found earlier by 
Feshbach and Tikochinsky \cite{FesTik}. 
In this way, the pseudo Bogoliubov transformation makes it possible to 
solve the eigenvalue problem of the Hamiltonian operator $H$ given in Eq. (\ref{2.12}).

Let us now consider the Schr\"{o}dinger equation 
\begin{align}
i\hbar \frac{d}{dt} |^{\s} \psi(t) \rangle =H |^{\s} \psi(t) \rangle \,. 
\label{3.17}
\end{align}
In order to solve this equation, we expand $|^{\s} \psi(t) \rangle$ in terms of 
the basis $\big\{^{\m} |^{\s} n_1, n_2 \dket \big\}$ at $t=0$ rather than 
the basis $\big\{^{\m} |^{\s} n_1, n_2 \ket \big\}$ at $t=0$.  
Then, using Eq. (\ref{3.15}), we obtain the particular solutions 
\begin{align}
\left|^{\s} \psi^{(\pm)}_{n_1,\m n_2}(t) \right\rangle 
:=\exp \left(-ih^{(\pm)}_{n_1,\m n_2} t/\hbar\right)  |^{\s} n_1, n_2 \dket_{\s t=0}  \,, 
\label{3.18}
\end{align}
which specify the Hamiltonian eigenstates at the time $t$. 
The general solution of Eq. (\ref{3.17}) is 
given by $|^{\s} \psi^{(\pm)}(t) \rangle 
=\sum_{n_1, \m n_2} c_{n_1, \m n_2} \big|^{\s} \psi^{(\pm)}_{n_1,\m n_2}(t) \big\rangle$, 
with $c_{n_1, \m n_2}$ being complex constants. 
We see that $\big|^{\s} \psi^{(-)}_{n_1,\m n_2}(t) \big\rangle$ represent decaying states  
while $\big|^{\s} \psi^{(+)}_{n_1,\m n_2}(t) \big\rangle$ represent growing states, 
regardless of the possible values of $n_{1}$ and $n_{2}$. 
This result is due to the presence of the constant term $\pm i\hbar \lambda$ 
that remains in $h^{(\pm)}_{n_1,\m n_2}$ even when ${n_{1}=n_{2}=0}$. 
Since $H$ is Hermitian with respect to the $\dagger$-conjugation, 
the dual Schr\"{o}dinger equation for $\langle \psi(t)|$ reads $d\langle \psi(t)|/dt =(-ih)^{-1} \langle \psi(t)|H$. 
Expanding $\langle \psi(t)|$ in terms of the dual basis $\big\{^{\s} \dbra n_1, n_2 |^{\m} \big\}$ at $t=0$, 
and using the eigenvalue equation 
$\dbra n_1, n_2 |^{\s}H^{(\pm)} =h^{(\pm)}_{n_1,\m n_2} \dbra n_1, n_2 |^{\s}$ obtained from Eq. (\ref{3.11b}),  
we have the particular solutions 
\begin{align}
\left\langle \psi^{(\pm)}_{n_1,\m n_2}(t) \s \right| 
:=\exp \left(ih^{(\pm)}_{n_1,\m n_2} t/\hbar\right)  \dbra n_1, n_2 |_{\m t=0} \,.  
\label{3.19}
\end{align}
It is clear, by taking into account the condition 
${\big(h^{(\pm)}_{n_1,\m n_2}\big)^{\ddagger}=h^{(\pm)}_{n_1,\m n_2}}$ ensured by Eq. (\ref{3.7}),  
that $\big|^{\s} \psi^{(\pm)}_{n_1,\m n_2}(t) \big\rangle$ 
and $\big\langle \psi^{(\pm)}_{n_1,\m n_2}(t)_{\s} \big|$ are related to each other by the $\ddagger$-conjugation. 
Equation (\ref{3.13}) leads to 
$\big\langle \psi^{(\pm)}_{m_1,\m m_2}(t)_{\s} \big|^{\s} \psi^{(\pm)}_{n_1,\m n_2}(t) \big\rangle 
=\delta_{m_1 n_1} \delta_{m_2 n_2}$,  
which demonstrates that the squared-norm of $\big|^{\s} \psi^{(\pm)}_{n_1,\m n_2}(t) \big\rangle$
does not change in time.  
A similar fact was also pointed out by Feshbach and Tikochinsky \cite{FesTik}.

Now, evaluating the standard squared-norm of $|^{\s} n_1, n_2 \dket$, 
we show that the ordinary Hilbert space is not well-defined in the cases ${\theta=\pm\pi/4}$.  
We first note that the vectors defined in Eq. (\ref{3.9}) can be written as 
\begin{subequations}
\label{3.20}
\begin{align}
|\s0\dket &=\frac{1}{\cos \theta} \exp\left( a_{1}^{\dagger} a_{2}^{\dagger} \tan\theta \right) |\s0\ket \,, 
\label{3.20a}
\\[2.5pt]
\dbra 0\s| &=\frac{1}{\cos \theta} \bra 0\s| \exp\left( -a_{1} a_{2} \tan\theta\s \right) , 
\label{3.20b}
\end{align}
\end{subequations}
or equivalently as 
\begin{subequations}
\label{3.201}
\begin{align}
|\s0\dket &=\frac{1}{\cos \theta} \sum_{n=0}^{\infty} \left(\tan\theta\right)^{n} |^{\s} n, n \ket \,, 
\label{3.201a}
\\[2.5pt]
\dbra 0\s| &=\frac{1}{\cos \theta} \sum_{n=0}^{\infty} \left(-\tan\theta\right)^{n} \bra n, n \s | \,. 
\label{3.201b}
\end{align}
\end{subequations}
These expressions are well-defined only for $\theta$ such that $|\tan\theta^{\,}|<1$, 
because the formula 
$\sum_{n=0}^{\infty} \big(-\tan^{2} \theta \big)^{n}=\big( 1+\tan^{2} \theta \big)^{-1}$ 
is used to prove that $\dbra 0^{\s}|^{\s}0\dket =1$. 
For this reason, the condition ${\theta=\pm\pi/4}$ should here be understood as 
${\theta\uparrow\pi/4}$ or ${\theta\downarrow -\pi/4}$. 
The $\dagger$-conjugate of $|^{\s}0\dket$ is defined from Eq. (\ref{3.9a}) as follows: 
\begin{align}
[\!\bra 0^{\s}|:=\bra 0\s|\s e^{\theta^{\ast} X^{\dagger}} =\bra 0\s|\s e^{\theta^{\ast} X} . 
\label{3.202} 
\end{align}
With the use of Eqs. (\ref{3.20a}) and (\ref{3.201a}), $[\!\bra 0^{\s}|$ can be expressed as 
\begin{subequations}
\label{3.203}
\begin{align}
[\!\bra 0^{\s}|&=\frac{1}{\cos \theta^{\ast}} \bra 0\s| \exp\left( a_{1} a_{2} \tan\theta^{\ast} \right) 
\label{3.203a}
\\[2.5pt]
&=\frac{1}{\cos \theta^{\ast}} \sum_{n=0}^{\infty} \left(\tan\theta^{\ast}\right)^{n} \bra n, n \s | \,.
\label{3.203b}
\end{align}
\end{subequations}
When $\theta$ is not purely imaginary, $[\!\bra 0^{\s}|$ is different from $\dbra 0^{\s}|$. 
From Eqs. (\ref{3.201a}) and (\ref{3.203b}),  
the standard squared-norm of $|^{\s}0\dket$ is found to be  
\begin{align}
[\!\bra 0^{\s} |\s0\dket =\frac{1}{|\cos\theta^{\,}|^{2}-|\sin\theta^{\,}|^{2}} 
=\frac{1}{\cos\left(\theta +\theta^{\ast} \right)} \,.
\label{3.21}
\end{align}
If $\theta$ is purely imaginary, then Eq. (\ref{3.21}) becomes ${\dbra 0^{\s}|\s0\dket}=1$ as expected. 
In contrast, if ${\theta=\pm\pi/4}$, then $[\!\bra 0^{\s} |\s0\dket$ diverges to infinity.

More generally, we can evaluate the standard squared-norm of $|^{\s} n_1, n_2 \dket$. 
Just like $[\!\bra 0^{\s}|$, the $\dagger$-conjugate of $|^{\s} n_1, n_2 \dket$ is defined 
from Eq. (\ref{3.12a}) by 
\begin{align}
[\!\bra  n_1, n_2 \s|:=\bra n_1, n_2 \s|^{\s} e^{\theta^{\ast} X^{\dagger}} =\bra n_1, n_2 \s|^{\s} e^{\theta^{\ast} X} . 
\label{3.210}
\end{align}
With this, the standard squared-norm of $|^{\s} n_1, n_2 \dket$ is given by 
\begin{align}
[\!\bra  n_1, n_2 \s|^{\s} n_1, n_2 \dket 
=\bra n_1, n_2 \s|^{\s} e^{(\theta+\theta^{\ast})X} |^{\s} n_1, n_2 \ket \,. 
\label{3.211}
\end{align}
For instance, we can obtain   
\begin{subequations}
\label{3.212}
\begin{align}
&[\!\bra 1, 0\s|^{\s} 1, 0 \dket =[\!\bra 0, 1\s|^{\s} 0, 1 \dket 
=\frac{1}{(\cos \varTheta)^{2}} \,,  
\label{3.212a}
\\
&[\!\bra 1, 1\s|^{\s} 1, 1 \dket =\frac{2-(\cos \varTheta)^{2}}{(\cos \varTheta)^{3}} \,,
\label{3.212b} 
\end{align}
\end{subequations}
where $\varTheta:=\theta +\theta^{\ast}$. 
Furthermore, the use of induction yields 
\begin{align}
[\!\bra  n_1, n_2 \s|^{\s} n_1, n_2 \dket 
\simeq \frac{c}{(\cos \varTheta)^{n_1 +n_2 +1}} \quad (\varTheta \rightarrow \pm \pi/2 ) \,, 
\label{3.213}
\end{align}
with $c$ being a real constant. 
Thus, it follows that since the standard squared-norm ${[\!\bra  n_1, n_2 \s|^{\s} n_1, n_2 \dket }$ 
diverges to infinity as ${\theta \rightarrow \pm\pi/4}$, 
the ordinary Hilbert space specified by the standard squared-norm 
is not well-defined in the cases ${\theta=\pm\pi/4}$.
Accordingly, it turns out that in the cases ${\theta=\pm\pi/4}$, 
the Hermiticity of $H_{1}$ does not actually make sense 
in the ordinary Hilbert space and, as a result, $H_{1}$ 
can possess the purely imaginary eigenvalues $\pm i\hbar \lambda {(n_1 +n_2 +1)}$ [see Eqs. (\ref{3.15}) and (\ref{3.16})]. 
This situation was also stated by Feshbach and Tikochinsky in a somewhat different context \cite{FesTik}. 
To avoid the use of the ill-defined Hilbert space mentioned here, 
we have indeed considered the well-defined Hilbert space specified by the inner product (\ref{3.13}). 
It is now clear that we should adopt this well-defined Hilbert space in the present quantization approach.

Using Eq. (\ref{2.4}) at the operator level and Eqs. (\ref{2.10}), (\ref{3.4}), and (\ref{3.8}), 
we can obtain for ${\theta=\pi/4}$, 
\begin{subequations}
\label{3.22}
\begin{align}
x(t)&=\sqrt{\frac{\hbar}{2m\omega}}\m e^{-\lambda t} 
\left(\s \bar{a}_{1}^{\ddagger}(0) \s e^{i\omega t} +\bar{a}_{2}^{\ddagger}(0) \s e^{-i\omega t} \s\right), 
\label{3.22a}
\\[3pt]
y(t)&=\sqrt{\frac{\hbar}{2m\omega}}\m e^{\lambda t} 
\left(\s \bar{a}_{1}(0) \s e^{-i\omega t} -\bar{a}_{2}(0) \s e^{i\omega t} \s\right), 
\label{3.22b}
\end{align}
\end{subequations}
and for $\theta=-\pi/4$, 
\begin{subequations}
\label{3.23}
\begin{align}
x(t)&=\sqrt{\frac{\hbar}{2m\omega}}\m e^{-\lambda t} 
\left(\s \bar{a}_{1}(0) \s e^{-i\omega t} +\bar{a}_{2}(0) \s e^{i\omega t} \s\right), 
\label{3.23a}
\\[3pt]
y(t)&=\sqrt{\frac{\hbar}{2m\omega}}\m e^{\lambda t} 
\left(\s \bar{a}_{1}^{\ddagger}(0) \s e^{i\omega t} -\bar{a}_{2}^{\ddagger}(0) \s e^{-i\omega t} \s\right) .
\label{3.23b}
\end{align}
\end{subequations}
It can be readily checked that Eqs. (\ref{3.22a}) and (\ref{3.23a}) satisfy Eq. (\ref{2.2}), 
and Eqs. (\ref{3.22b}) and (\ref{3.23b}) satisfy Eq. (\ref{2.3}).  
We thus see that Eqs. (\ref{2.2}) and (\ref{2.3}) at the operator level are realized 
in Feshbach-Tikochinsky's quantization approach reformulated here.

We close this section with a remark on the Hamiltonian operator (\ref{2.12}). 
This operator has the same form as 
one of the Hamiltonian operators argued in TFD \cite{TakUme, Umezawa}, 
provided that $a_{2}$ and $a_{2}^{\dagger}$ are identified with the so-called {\em tilde conjugates} of  
$a_{1}$ and $a_{1}^{\dagger}$, respectively. 
Noting this fact, Celeghini {\it et al.} have investigated 
a quantum-theoretical aspect of the Bateman model 
by following the framework of TFD \cite{CRV, SVW}. 
They claimed the necessity of a field theoretical generalization of the Bateman model. 
In the thermo field dynamical approach, 
the minus sign of ${-}^{\s}a_{2}^{\dagger} a_{2}$ included in $H_{0}$ is essential for 
describing the thermal degree of freedom. 
However, from a purely dynamical point of view, this minus sign inevitably causes the problem 
of dynamical instability of the system if interactions are turned on. 
In fact, the eigenvalues of $H_{0}$, which are given by 
$\mathrm{Re}^{\s} h^{(\pm)}_{n_1,\m n_2}=\hbar \omega {(n_1 -n_2)}$ and 
can be identified as the possible values of energy,  
are unbounded from below owing to the presence of ${-}^{\s}n_2$. 
As a result, the dynamical stability of the system is spoiled. 
This undesirable situation can be overcome by applying the imaginary-scaling quantization 
scheme \cite{BenMan, Mostafazadeh} to the Bateman model.

\section{Imaginary-scaling quantization approach}

In this section, we treat the imaginary-scaling quantization of the Bateman model.

First we define the operators $\tilde{a}_{i}$ and $\tilde{a}_{i}^{\S}$ by
\begin{align}
\tilde{a}_{i}:=e^{\phi Y} a_{i} \m e^{-\phi Y} \s,
\quad\;\,
\tilde{a}_{i}^{\S}:=e^{\phi Y} a_{i}^{\dagger} e^{-\phi Y} \s,
\label{4.1}
\end{align}
where $\phi$ is a complex parameter, and $Y$ is defined by
\begin{align}
Y:=-\frac{i}{2} \left( a_{2}^{2}-a_{2}^{\dagger\s 2} \right) . 
\label{4.2}
\end{align}
It is obvious that $Y^{\dagger}=Y$. The unitarity of $e^{\phi Y}$ and its associated property 
$(\tilde{a}_{i})^{\dagger}=\tilde{a}_{i}^{\S}$ hold only when $\phi$ is purely imaginary. 
We can express $Y$ as 
$Y=-i^{\s}\big( \tilde{a}_{2}^{2}-\tilde{a}_{2}^{\S}{}^{2} \big)/2$, from which we see that 
$Y$ is Hermitian with respect to the $\S$-conjugation, i.e., $Y^{\s\S}=Y$. 
Equation ({2.11}) leads to 
\begin{align}
\left[\s \tilde{a}_{i}\s, \s\tilde{a}_{j}^{\S} \s \right]\!=\delta_{ij} \1 \,, 
\quad\; \mbox{all others}=0 \,. 
\label{4.3}
\end{align}
From the definition of $Y$, we immediately see that 
\begin{align}
\tilde{a}_{1}=a_{1} \,,  \quad\;\, \tilde{a}_{1}^{\S}=a_{1}^{\dagger} \,. 
\label{4.4}
\end{align}
Thus it turns out that the transformation 
$\big(a_{i}, a_{i}^{\dagger}\big) \rightarrow \big(\tilde{a}_{i}, \tilde{a}_{i}^{\S} \big)$ 
is essentially a squeeze transformation of $\big(a_{2}, a_{2}^{\dagger}\big)$, 
provided that $\phi$ is purely imaginary and hence $e^{\phi Y}$ is unitary \cite{Stoler, Umezawa}. 
From now on, we rather choose $\phi$ to be the real value ${\phi=\pi/2}$. Then, from Eq. (\ref{4.1}), we have  
\begin{align}
\tilde{a}_{2}=-ia_{2}^{\dagger} \,,  \quad\;\, \tilde{a}_{2}^{\S}=-ia_{2} \,. 
\label{4.5}
\end{align}
The transformation 
$\big(a_{2}, a_{2}^{\dagger}\big) \rightarrow \big(\tilde{a}_{2}, \tilde{a}_{2}^{\S} \big)
={\big(-ia_{2}^{\dagger}, -ia_{2}\big)}$ 
is precisely the imaginary-scaling transformation argued in Refs. \cite{BenMan, Mostafazadeh}. 
Since this transformation can be derived as a non-unitary analog of the squeeze transformation, 
it can be said to be a {\em pseudo} squeeze transformation.

Next we define the operators $\check{a}_{i}$ and $\check{a}_{i}^{\S}$ by 
\begin{align}
\check{a}_{i}:=e^{\,\chi Z} \tilde{a}_{i} \m e^{-\chi Z} \s,
\quad\;\,
\check{a}_{i}^{\S}:=e^{\,\chi Z} \tilde{a}_{i}^{\S} e^{-\chi Z} \s,
\label{4.6}
\end{align}
where $\chi$ is assumed to be a purely imaginary parameter 
satisfying $\chi^{\S}=-\chi$, 
and $Z$ is defined by 
\begin{align}
Z:=\tilde{a}_{1}^{\S} \tilde{a}_{2} +\tilde{a}_{2}^{\S} \tilde{a}_{1} \,.
\label{4.7}
\end{align}
Obviously, $Z$ is Hermitian with respect to the $\S$-conjugation. 
The unitarity of $e^{\,\chi Z}$ with respect to the $\S$-conjugation and 
the $\S$-conjugation relation $\big(\check{a}_{i} \big)^{\S}=\check{a}_{i}^{\S}$ are ensured accordingly. 
Equation (\ref{4.3}) leads to 
\begin{align}
\left[\s \check{a}_{i}\s, \s\check{a}_{j}^{\S} \s \right]\!=\delta_{ij} \1 \,, 
\quad\; \mbox{all others}=0 \,. 
\label{4.8}
\end{align}
The operators $\check{a}_{i}$ can be written as linear combinations of $\tilde{a}_{1}$ and $\tilde{a}_{2}$; 
similarly, the operators $\check{a}_{i}^{\S}$ can be written as linear combinations of 
$\tilde{a}_{1}^{\S}$ and $\tilde{a}_{2}^{\S}$. 
The transformation 
$\big(\tilde{a}_{i}, \tilde{a}_{i}^{\S}\big) \rightarrow \big(\check{a}_{i}, \check{a}_{i}^{\S} \big)$ 
is thus realized as a homogeneous  transformation. 
Combining the expressions of the linear combinations with Eqs. (\ref{4.4}) and (\ref{4.5}), we obtain 
\begin{subequations}
\label{4.9}
\begin{align}
\left( 
\begin{array}{c}
\check{a}_{1} \\[1pt]
\check{a}_{2} \\ 
\end{array}
\right)
&=
\left(
\begin{array}{cc}
\cosh\chi &\, i\sinh\chi \\[1pt]
-\sinh\chi &\, -i\cosh\chi \\
\end{array}
\right) 
\left( 
\begin{array}{c}
a_{1} \\[1pt]
a_{2}^{\dagger} \\ 
\end{array}
\right) \s,
\label{4.9a}
\\[5pt]
\left( 
\begin{array}{c}
\check{a}_{1}^{\S} \\[1pt]
\check{a}_{2}^{\S} \\ 
\end{array}
\right)
&=
\left(
\begin{array}{cc}
\cosh\chi &\, -i\sinh\chi \\[1pt]
\sinh\chi &\, -i\cosh\chi \\
\end{array}
\right) 
\left( 
\begin{array}{c}
a_{1}^{\dagger} \\[1pt]
a_{2} \\ 
\end{array}
\right) \s.
\label{4.9b}
\end{align}
\end{subequations}

Now, using Eqs. (\ref{4.9}) and (\ref{4.8}), we can express the operators $H_{0}$ and $H_{1}$  
defined in Eq. (\ref{2.13}) as follows:\footnote{~The symmetry group associated with 
the imaginary-scaling quantization approach is ${\mathit{SU}(2)}$, instead of ${\mathit{SU}(1,1)}$, 
as can be seen from Eq. (\ref{4.10a}).  
We note the fact that $H_{0}$ in Eq. (\ref{2.13a}) corresponds to the Casimir operator of ${\mathit{SU}(1,1)}$, 
while $H_{0}$ in Eq. (\ref{4.10a}) corresponds to the Casimir operator of ${\mathit{SU}(2)}$.
As can be easily seen, ${\mathit{SU}(2)}$ is related to ${\mathit{SU}(1,1)}$ by 
the imaginary-scaling transformation 
$\big(a_{2}, a_{2}^{\dagger}\big) \rightarrow \big(\tilde{a}_{2}, \tilde{a}_{2}^{\S} \big)
={\big(-ia_{2}^{\dagger}, -ia_{2}\big)}$.}
\begin{subequations}
\label{4.10}
\begin{align}
H_{0} &=\hbar \omega \left( \check{a}_{1}^{\S} \check{a}_{1} +\check{a}_{2}^{\S} \check{a}_{2} +\1 \right) ,
\label{4.10a}
\\[2pt]
H_{1} &=\frac{\hbar\gamma}{2m}  
\left\{ \left( \check{a}_{1}^{\S} \check{a}_{2} -\check{a}_{2}^{\S} \check{a}_{1} \right) \cosh(2\chi) 
+\left( \check{a}_{1}^{\S} \check{a}_{1} -\check{a}_{2}^{\S} \check{a}_{2} \right) \sinh(2\chi) \right\} . 
\label{4.10b}
\end{align}
\end{subequations}
Incidentally, $Z$ can be expressed as 
$Z=\check{a}_{1}^{\S} \check{a}_{2} +\check{a}_{2}^{\S} \check{a}_{1}$. 
Our present purpose is to find the eigenvalues of $H=H_{0}+H_{1}$ 
within the framework of imaginary-scaling quantization. 
To this end, we choose $\chi$ to be the imaginary value ${\chi=\pm i\pi /4}$  
so that $H_{1}$ can take the form of a linear combination of  
$\check{a}_{1}^{\S} \check{a}_{1}$ and $\check{a}_{2}^{\S} \check{a}_{2}$. 
(The operator $H_{0}$ already takes the form of a  linear combination of  
$\check{a}_{1}^{\S} \check{a}_{1}$,  $\check{a}_{2}^{\S} \check{a}_{2}$, and $\1$.)  
After setting ${\chi=\pm i\pi/4}$, the operator $H_{1}$ reduces to   
\begin{align}
\check{H}_{1}^{(\pm)} &:=\pm i \frac{\hbar\gamma}{2m}  
\left( \check{a}_{1}^{\S} \check{a}_{1} -\check{a}_{2}^{\S} \check{a}_{2} \right) . 
\label{4.11}
\end{align}
The Hermiticity of $\check{H}_{1}^{(\pm)}$ with respect to the $\S$-conjugation, i.e., 
$\big(\check{H}_{1}^{(\pm)} \big)^{\S}=\check{H}_{1}^{(\pm)}$ is valid under the conditions 
\begin{align}
i^{\s\S}=-i\,, \quad\;\, \gamma^{\S}=-\gamma \,.
\label{4.12}
\end{align}
It is obvious that $H_{0}$ and $Z$ are Hermitian with respect to the $\S$-conjugation.

The Hamiltonian operator (\ref{2.12}) now reads $\check{H}^{(\pm)}=H_{0}+\check{H}_{1}^{(\pm)}$. 
Correspondingly, the Heisenberg equation for an implicitly time-dependent operator $A(t)$ reads 
$dA/dt=(i\hbar)^{-1} \big[^{\s}A, \check{H}^{(\pm)} \big]$. 
By using the commutation relations in Eq. (\ref{4.8}), 
we can solve the Heisenberg equations for $\check{a}_{i}$ and $\check{a}_{i}^{\S}$, obtaining     
\begin{subequations}
\label{4.13}
\begin{alignat}{3}
\check{a}_{1}(t) &=\check{a}_{1}(0)\s e^{(-i\omega \pm \lambda)\s t} \,, & \quad\;\, 
\check{a}_{1}^{\S}(t) &=\check{a}_{1}^{\S}(0) \s e^{-(-i\omega \pm \lambda)\s t} \,,
\label{4.13a}
\\[2pt]
\check{a}_{2}(t) &=\check{a}_{2}(0)\s e^{(-i\omega \mp \lambda)\s t} \,, & \quad\;\,
\check{a}_{2}^{\S}(t) &=\check{a}_{2}^{\S}(0)\s e^{-(-i\omega \mp \lambda)\s t} \,, 
\label{4.13b}
\end{alignat}
\end{subequations} 
with ${\lambda:=\gamma/2m}$. 
By virtue of the conditions in Eq. (\ref{4.12}), 
the $\S$-conjugation relation  
$\big(\check{a}_{i}(t)\big)^{\S}=\check{a}_{i}^{\S}(t)$ holds at an arbitrary time. 
We see from Eq. (\ref{4.13}) that 
just like the $\ddagger$-conjugation treated in Sec. 3, 
the $\S$-conjugation also involves time reversal. 
It is evident that the Hamiltonian operator $\check{H}^{(\pm)}$ is independent of time.

Let us define the new vectors 
\begin{align}
|\s0\dcket :=e^{\phi Y} |\s0\ket \,, 
\quad\;\,
\dcbra 0\s|:=\bra 0\s|\s e^{-\phi Y} \s, 
\label{4.14}
\end{align}
which satisfy 
\begin{align}
\tilde{a}_{i}\s |\s0\dcket=0 \,, 
\quad\;\,
\dcbra 0\s|\s \tilde{a}_{i}^{\S}=0 
\label{4.15}
\end{align}
owing to Eqs. (\ref{2.14}) and (\ref{2.16}).\footnote{~If $|\tan\phi^{\,}|<1$, 
the vectors $|\s0\dcket$ and $\dcbra 0\s|$ can be written as
\begin{align*}
|\s0\dcket &=\frac{1}{\sqrt{\cos \phi}} \exp\left( \s\frac{i}{2} a_{2}^{\dagger\s 2} \tan\phi \right) |\s0\ket \,, 
\\
\dcbra 0\s| &=\frac{1}{\sqrt{\cos \phi}} \bra 0\s| \exp\left( \s\frac{i}{2} a_{2}^{2} \tan\phi\s \right) .
\end{align*}
These expressions cannot be applied to the present case, $\phi=\pi/2$. 
}
Also, using Eqs. (\ref{4.6}) and (\ref{4.15}), we can show that  
\begin{align}
\check{a}_{i}\s |\s0\dcket=0 \,, 
\quad\;\,
\dcbra 0\s|\s \check{a}_{i}^{\S}=0 \,.
\label{4.16}
\end{align}
Hence, $|\s0\dcket$ and $\dcbra 0\s|$ are established as the vacuum state vectors 
common to both the 
$\big(\tilde{a}_{i}, \tilde{a}_{i}^{\S} \big)$ and $\big(\check{a}_{i}, \check{a}_{i}^{\S} \big)$
systems. 
From Eqs. (\ref{4.15}) and (\ref{4.16}), it turns out that $\tilde{a}_{i}$ and $\check{a}_{i}$
are annihilation operators, while $\tilde{a}_{i}^{\S}$ and $\check{a}_{i}^{\S}$
are creation operators. 
In the $\big(\check{a}_{i}, \check{a}_{i}^{\S} \big)$-system, 
we can construct the Fock basis vectors and their dual vectors as follows: 
\begin{subequations}
\label{4.17}
\begin{align}
|^{\s} n_1, n_2 \dcket &:=\frac{1}{\sqrt{n_{1}! \, n_{2}!}} 
\Big(\check{a}_{1}^{\S} \Big)^{n_1} \Big(\check{a}_{2}^{\S} \Big)^{n_2} |\s0\dcket \m, 
\label{4.17a}
\\[2.5pt]
\dcbra n_1, n_2 | &:=\frac{1}{\sqrt{n_{1}! \, n_{2}!}} 
\dcbra 0\s|(\check{a}_{1})^{n_1} (\check{a}_{2})^{n_2} \s. 
\label{4.17b}
\end{align}
\end{subequations}
They are related to the old basis vectors in Eqs. (\ref{2.15}) and (\ref{2.17}) by 
\begin{subequations}
\label{4.18}
\begin{align}
|^{\s} n_1, n_2 \dcket &=e^{\chi Z } e^{\phi Y} |^{\s} n_1, n_2 \ket \,, 
\label{4.18a}
\\[2pt]
\dcbra n_1, n_2 | &=\bra n_1, n_2 |\s e^{-\phi Y} e^{-\chi Z} \s. 
\label{4.18b}
\end{align}
\end{subequations}
It is easy to show by using Eq. (\ref{2.18}) that 
\begin{align}
\dcbra m_1, m_2 \s |^{\s} n_1, n_2 \dcket=\delta_{m_1 n_1} \delta_{m_2 n_2} \m.
\label{4.19}
\end{align}
From this, it follows that the Fock basis vectors $|^{\s} n_1, n_2 \dcket$ have the positive squared-norm 1, and 
the Fock space spanned by the orthonormal basis $\big\{^{\m} |^{\s} n_1, n_2 \dcket \big\}$ 
is a positive-definite Hilbert space. 
The completeness relation (\ref{2.19}) now leads to 
\begin{align}
\sum_{n_1=0}^{\infty} \sum_{n_2=0}^{\infty}  |^{\s} n_1, n_2 \dcket \dcbra n_1, n_2 \s |=\1 \,.
\label{4.20}
\end{align}
We immediately see that 
the vectors  
$|^{\s} n_1, n_2 \dcket$ with ${\phi=\pi/2}$ and ${\chi=\pm i\pi /4}$ 
are simultaneous eigenvectors of 
$H_{0}$ and $\check{H}_{1}^{(\pm)}$ and satisfy the Hamiltonian eigenvalue equation 
\begin{align}
\check{H}^{(\pm)} |^{\s} n_1, n_2 \dcket =\check{h}^{(\pm)}_{n_1,\m n_2} |^{\s} n_1, n_2 \dcket 
\label{4.21}
\end{align}
with   
\begin{align}
\check{h}^{(\pm)}_{n_1,\m n_2} :=\hbar \omega (n_1 +n_2 +1) 
\pm i\hbar \lambda (n_1 -n_2) \,.
\label{4.22}
\end{align}
This expression of the Hamiltonian eigenvalues is completely different from the one obtained by 
Feshbach and Tikochinsky, namely Eq. (\ref{3.16}). 
In fact, the eigenvalues of $H_{0}$, which are given at present by 
$\mathrm{Re}^{\s} \check{h}^{(\pm)}_{n_1,\m n_2}=\hbar \omega {(n_1 +n_2 +1)}$, 
are bounded from below, and therefore the dynamical stability of the system is ensured.  
Also, $\mathrm{Re}^{\s} \check{h}^{(\pm)}_{n_1,\m n_2}$ include the vacuum state energy $\hbar\omega$. 
In this way, 
the combination of the imaginary-scaling transformation and a homogeneous transformation 
makes it possible to solve the eigenvalue problem of the Hamiltonian operator $H$ given in Eq. (\ref{2.12}),  
resolving the problem of dynamical instability encountered in Feshbach-Tikochinsky's quantization approach.

Now we recall the Schr\"{o}dinger equation (\ref{3.17}) and expand the state vector $|^{\s} \psi(t) \rangle$ 
in terms of the basis $\big\{^{\m} |^{\s} n_1, n_2 \dcket \big\}$ at $t=0$, instead of 
the basis $\big\{^{\m} |^{\s} n_1, n_2 \dket \big\}$ at $t=0$. 
Then, using Eq. (\ref{4.21}), we obtain the particular solutions of the Schr\"{o}dinger equation, 
\begin{align}
\left|^{\s} \check{\psi}{}^{(\pm)}_{n_1,\m n_2}(t) \right\rangle 
:=\exp \left(-i\check{h}^{(\pm)}_{n_1,\m n_2} t/\hbar\right)  |^{\s} n_1, n_2 \dcket_{\s t=0} \,,   
\label{4.23}
\end{align}
which specify the Hamiltonian eigenstates at the time $t$. 
The general solution is found to be $|^{\s} \check{\psi}^{(\pm)}(t) \rangle 
=\sum_{n_1, \m n_2} \check{c}_{n_1, \m n_2} \big|^{\s} \check{\psi}{}^{(\pm)}_{n_1,\m n_2}(t) \big\rangle$, 
with $\check{c}_{n_1, \m n_2}$ being complex constants. 
We see that both $\big|^{\s} \check{\psi}{}^{(+)}_{n_1,\m n_2}(t) \big\rangle$ and 
$\big|^{\s} \check{\psi}{}^{(-)}_{n_1,\m n_2}(t) \big\rangle$ can represent either of decaying or growing states   
depending on the possible values of $n_{1}$ and $n_{2}$. 
If $\big|^{\s} \check{\psi}{}^{(+)}_{n_1,\m n_2}(t) \big\rangle$ is the state vector of a decaying (growing) state, 
then $\big|^{\s} \check{\psi}{}^{(-)}_{n_1,\m n_2}(t) \big\rangle$ is the state vector of a growing (decaying) state. 
It is remarkable that the state vectors $\big|^{\s} \check{\psi}{}^{(\pm)}_{n_1,\m n_2}(t) \big\rangle$ 
with ${n_{1}=n_{2}}$ contain no $\gamma$ and 
represent stable states, because $\mathrm{Im}^{\s} \check{h}^{(\pm)}_{n_1,\m n_2}$ vanish when ${n_{1}=n_{2}}$.  
Therefore, unlike Feshbach-Tikochinsky's quantization approach, 
the imaginary-scaling quantization approach allows to have stable states in addition to 
decaying states and growing states. 
Recall here the dual Schr\"{o}dinger equation $d\langle \psi(t)|/dt =(-ih)^{-1} \langle \psi(t)|H$. 
Expanding $\langle \psi(t)|$ in terms of the dual basis $\big\{^{\s} \dcbra n_1, n_2 |^{\m} \big\}$ at $t=0$, 
and using the eigenvalue equation 
$\dcbra n_1, n_2 |^{\s} \check{H}{}^{(\pm)} =\check{h}^{(\pm)}_{n_1,\m n_2} \dcbra n_1, n_2 |^{\s}$ obtained from Eq. (\ref{4.17b}),  
we have the particular solutions 
\begin{align}
\left\langle \check{\psi}{}^{(\pm)}_{n_1,\m n_2}(t) \s \right| 
:=\exp \left(i \check{h}^{(\pm)}_{n_1,\m n_2} t/\hbar\right)  \dcbra n_1, n_2 |_{\m t=0} \,.  
\label{4.24}
\end{align}
Taking into account the condition 
$\big(\check{h}^{(\pm)}_{n_1,\m n_2}\big)^{\S}=\check{h}^{(\pm)}_{n_1,\m n_2}$ ensured by Eq. (\ref{4.12}),  
we see that $\big|^{\s} \check{\psi}{}^{(\pm)}_{n_1,\m n_2}(t) \big\rangle$ 
and $\big\langle \check{\psi}{}^{(\pm)}_{n_1,\m n_2}(t)_{\s} \big|$ are related to each other by the $\S$-conjugation. 
Equation (\ref{4.19}) leads to 
$\big\langle \check{\psi}{}^{(\pm)}_{m_1,\m m_2}(t)_{\s} \big|^{\s} \check{\psi}{}^{(\pm)}_{n_1,\m n_2}(t) \big\rangle 
=\delta_{m_1 n_1} \delta_{m_2 n_2}$,  
which implies that the squared-norm of $\big|^{\s} \check{\psi}{}^{(\pm)}_{n_1,\m n_2}(t) \big\rangle$
does not change in time. A similar result was also found in Sec 3.

Using Eq. (\ref{2.4}) at the operator level and Eqs. (\ref{2.10}), (\ref{4.9}), and (\ref{4.13}), 
we can obtain for $\chi=i\pi/4$, 
\begin{subequations}
\label{4.25}
\begin{align}
x(t)&=\sqrt{\frac{\hbar}{2m\omega}}\m e^{-\lambda t} 
\left(\s \check{a}_{1}^{\S}(0) \s e^{i\omega t} +i\check{a}_{2}(0) \s e^{-i\omega t} \s\right), 
\label{4.25a}
\\[3pt]
y(t)&=\sqrt{\frac{\hbar}{2m\omega}}\m e^{\lambda t} 
\left(\s \check{a}_{1}(0) \s e^{-i\omega t} -i\check{a}_{2}^{\S}(0) \s e^{i\omega t} \s\right), 
\label{4.25b}
\end{align}
\end{subequations}
and for ${\chi=-i\pi/4}$, 
\begin{subequations}
\label{4.26}
\begin{align}
x(t)&=\sqrt{\frac{\hbar}{2m\omega}}\m e^{-\lambda t} 
\left(\s \check{a}_{1}(0) \s e^{-i\omega t} +i\check{a}_{2}^{\S} (0) \s e^{i\omega t} \s\right), 
\label{4.26a}
\\[3pt]
y(t)&=\sqrt{\frac{\hbar}{2m\omega}}\m e^{\lambda t} 
\left(\s \check{a}_{1}^{\S}(0) \s e^{i\omega t} -i\check{a}_{2} (0) \s e^{-i\omega t} \s\right) .
\label{4.26b}
\end{align}
\end{subequations}
It can be immediately checked that Eqs. (\ref{4.25a}) and (\ref{4.26a}) satisfy Eq. (\ref{2.2}), 
and Eqs. (\ref{4.25b}) and (\ref{4.26b}) satisfy Eq. (\ref{2.3}).  
In this way, it is verified that Eqs. (\ref{2.2}) and (\ref{2.3}) at the operator level are realized 
also in the imaginary-scaling quantization approach. 
In each of the cases ${\chi=i\pi/4}$ and ${\chi=-i\pi/4}$, we observe that 
$x^{\S}=y$ and $y^{\S}=x$. 
From these relations together with $\gamma^{\S}=-\gamma$ given in Eq. (\ref{4.12}), 
we can recognize that the $\S$-conjugation corresponds to the transformation 
$(x, y, \gamma) \rightarrow (^{\s}y, x, -\gamma)$, which leaves the Lagrangian (\ref{2.1}) invariant. 
On the other hand, it is recognized in Feshbach-Tikochinsky's quantization approach 
that the $\ddagger$-conjugation corresponds to the transformation 
$(x, y, \gamma) \rightarrow (^{\s} \pm ip_{x}/m\omega, \mp ip_{y}/m\omega, -\gamma)$, 
where $p_{x}$ and $p_{y}$ denote the canonical momenta conjugate to $x$ and $y$, respectively. 
From this result, we see that unlike the $\S$-conjugation, the $\ddagger$-conjugation 
does not have its classical counterpart at the Lagrangian level.

\section{Summary and discussion}

We have investigated two quantization approaches to the Bateman model. 
One is Feshbach-Tikochinsky's quantization approach reformulated concisely 
without invoking the ${\mathit{SU}(1,1)}$ Lie algebra, 
and the other is the imaginary-scaling quantization approach proposed originally for the Pais-Uhlenbeck model. 
The former has been developed by applying a pseudo Bogoliubov transformation to the Bateman model, 
while the latter has been developed by applying 
the imaginary-scaling transformation and a homogeneous transformation to the Bateman model.  
The two quantization approaches can thus be realized on an equal footing 
on the basis of the different transformations of $a_{i}$ and $a_{i}^{\dagger}$. 
Also, we have pointed out that the imaginary-scaling transformation can be said to be 
a pseudo squeeze transformation.

We have indeed solved the eigenvalue problem for the Hamiltonian operator $H$ of the Bateman model. 
By means of the pseudo Bogoliubov transformation, we have simply derived the Hamiltonian eigenvalues 
$h^{(\pm)}_{n_1,\m n_2}$ that were found earlier by Feshbach and Tikochinsky \cite{FesTik}. 
In addition, we have derived the alternative Hamiltonian eigenvalues $\check{h}^{(\pm)}_{n_1,\m n_2}$ by 
employing the imaginary-scaling quantization scheme \cite{BenMan, Mostafazadeh}. 
It has been seen that the real part of $h^{(\pm)}_{n_1,\m n_2}$ is proportional to ${n_{1}-n_{2}}$   
and the imaginary part is proportional to ${n_{1}+n_{2}+1}$. 
In contrast, the real part of $\check{h}^{(\pm)}_{n_1,\m n_2}$ is proportional to ${n_{1}+n_{2}+1}$   
and the imaginary part is proportional to ${n_{1}-n_{2}}$. 
As has been clarified above, the eigenvalues $\check{h}^{(\pm)}_{n_1,\m n_2}$ are desirable 
than $h^{(\pm)}_{n_1,\m n_2}$ from a purely dynamical point of view because 
$\mathrm{Re}^{\s} \check{h}^{(\pm)}_{n_1,\m n_2}$ are bounded from below. 
(By contrast, $h^{(\pm)}_{n_1,\m n_2}$ is desirable from the point of view of TFD.) 
With $\check{h}^{(\pm)}_{n_1,\m n_2}$, we have obtained the particular solutions of 
the Schr\"{o}dinger equation as in Eq. (\ref{4.23}). Then we have pointed out that 
the particular solutions with ${n_{1}=n_{2}}$ represent stable states. 
Such states do not appear in Feshbach-Tikochinsky's quantization approach. 
Also, the stable states cannot be understood at the classical mechanical level, because 
all the solutions of Eq. (\ref{2.2}) represent damped oscillations  
and all the solutions of Eq. (\ref{2.3}) represent amplified oscillations, provided that 
${4mk>\gamma^{2}}$. 
The emergence of the stable states might be viewed as a stabilization of the Bateman model 
occurring at the quantum-mechanical level.

We have been able to obtain the two different sets of eigenvalues 
$\big\{ h^{(\pm)}_{n_1,\m n_2} \big\}$ and $\big\{ \check{h}^{(\pm)}_{n_1,\m n_2} \big\}$ 
that correspond, respectively, to the two unitary inequivalent basis 
$\big\{^{\m} |^{\s} n_1, n_2 \dket \big\}$ and $\big\{^{\m} |^{\s} n_1, n_2 \dcket \big\}$   
determined for the one operator $H$. 
From this fact, we see that quantum mechanics has, so to speak, 
flexibility in deriving the set of possible values of a dynamical variable such as $H$. 
That is, the set of possible values of a dynamical variable is obtained depending on the choice of basis. 
This flexibility originates in the fact that quantum mechanics is composed of 
two basic objects -- dynamical variables (treated as operators) and state vectors,  
differently from classical mechanics, which is composed only of dynamical variables.

From a physical point of view, it would be interesting to find out dispersive systems 
to which our present formulations and results are applicable. 
In addition, the extensions of the Bateman model to interacting systems and many body systems 
remain as an interesting challenge that leads to field theoretical extensions of the Bateman model.

The Bateman model treats both the damped and amplified harmonic oscillators 
simultaneously on even ground, and therefore cannot be said to be a model only for 
the damped harmonic oscillator. 
To consistently treat only the damped harmonic oscillator within the framework of analytical mechanics, 
we need to find a new Lagrangian that, unlike 
the Caldirola-Kanai Lagrangian,\footnote{~The so-called Caldirola-Kanai Lagrangian reads
\cite{Razavy} 
\begin{align*}
L_{\rm CK}=e^{(\gamma/m) t} \left(\frac{m}{2} \dot{x}^{2} -\frac{k}{2} x^{2} \right) ,
\end{align*}
which indeed yields Eq. (\ref{2.2}) and describes only the damped harmonic oscillator. 
However, it has been pointed out that 
the canonical quantization based on $L_{\rm CK}$ is accompanied by some problems \cite{Brittin, Greenberger}. 
The Hamiltonian corresponding to $L_{\rm CK}$, rather than $L_{\rm CK}$ itself, was actually considered  
by Caldirola and Kanai independently \cite{Caldirola, Kanai, Razavy}. 
The doubled Lagrangian $2L_{\rm CK}$ was earlier found by Bateman \cite{Bateman} by substituting $y=e^{(\gamma/m) t} x$ 
into the Bateman Lagrangian (\ref{2.1}). For this reason, $L_{\rm CK}$ is 
sometimes called the Bateman-Caldirola-Kanai Lagrangian. 
The Lagrangian $L_{\rm CK}$ can be realized as a special case of the standard Lagrangian 
treated, e.g., in Ref. \cite{CieNik}. 
}
does not explicitly depend on time. 
This issue should be addressed in the near future.

\end{document}